\documentclass[10pt,amsmath,amssymb,prl]{revtex4}
\usepackage{graphicx}

\begin{document}
\title{Optimal protocol for quantum state tomography}
\author {E. V. Moreva}
\email{ekaterina.moreva@gmail.com} \affiliation{Moscow Engineering
Physics Institute (State University) Russia}
\author{Yu. I. Bogdanov, A.K. Gavrichenko}
\affiliation{Institute of Physics and Technology, Russian Academy of
Science}
\author{S.P. Kulik, I. Tikhonov} \affiliation{Faculty of Physics, Moscow State
University, 119992, Moscow, Russia}

\begin{abstract}
\noindent We develop a practical quantum tomography protocol and
implement measurements of pure states of ququarts realized with
polarization states of photon pairs (biphotons). The method is based
on an optimal choice of the measuring scheme's parameters that
provides better quality of reconstruction for the fixed set of statistical
data. A high accuracy of the state reconstruction (above
0.99) indicates that developed methodology is adequate.
\end{abstract}
\maketitle
\section{Introduction.}
For the last several years many elegant experiments were performed
in which different kinds of multi-dimensional quantum states
(qudits) were introduced \cite{qudits} (for more details see the
review \cite{Genovese07}). Most of them are based on the states of
light emitted via spontaneous parametric down-conversion (SPDC). In
this process photons of the laser pump decay on the pairs of photons
(biphotons) inside the crystal possessing non-zero quadratic
susceptibility. In stationary case the sum of daughter photons
frequencies coincides with the frequency of the pump
$w_{s}+w_{i}=w_{p}$ and intensity of the biphoton light emitted from
the crystal is maximal when phase matching condition holds
$k_{s}+k_{i}=k_{p}$. Choosing particular regime of SPDC one can
prepare a whole family of the biphoton states with different
properties. In this paper we focus upon polarization states of
biphotons although some other qudits can be realized with photon
pairs. Selecting so called frequency degenerate ($w_{s}=w_{i}$) and
collinear ($k_{s}||k_{i}$) regime of SPDC, polarization qutrits
(d=3) can be realized. Preparation of arbitrary polarization qutrit
(d=3) was reported in \cite{ourPRL:04}. The experimental method for
engineering with pure states of ququarts (d=4) was presented in \cite{ourPRA:06, D'Ariano}.\\
Quality assurance of preparation and transformation requires a
complete characterization of these states, which can be accomplished
through a procedure known as quantum process tomography. The first
protocol for reconstruction of polarization qutrits was introduced in
\cite{JetpLett:03,Jetp:03}. The protocol for quantum tomography of
polarization states of photon pair propagating along two spatial modes
was suggested in \cite{Kwiat:01}. Further this protocol was implemented
and modified for the case of single spatial mode \cite{ourPRA:06}.
However in order to distinguish the photons forming biphoton,
non-degenerate regime of SPDC was accomplished to realize a ququart.
This sort of states seems to be promising for quantum communication problems
because it allows one to pass the states of two photons (both entangled and product)
along single spatial mode for example in optical fiber.\\
Although tomographic procedure has been used in many experiments and right now serves as an
"application tool", still there are several problems related to simple
and in some sense optimal choice of protocol. Speaking about "optimal" protocol
we mean firstly how to implement the procedure with minimal number of measurements
for achieving highest accuracy. For example the remarkable paper \cite{Rehacek}
analyzes a minimal measurement scheme for single-qubit tomography.
Their analysis showed that the scheme is efficient in the sense that
it enables one to estimate the qubit state without enormous number
of qubits - a few thousand are sufficient for most practical
applications. Also the work \cite{Rehacek} indicates algorithms for manipulations
with qudits.  But from practical point of view there is the second reason to introduce
an optimality of the protocol. Usually the experimentalist possess by limited resources
to perform the measurements. For example he has a set of retardant plates with fixed
optical thickness but he can not access any other particular plates which are
necessary to perform the protocol according to an optimal way (see previous point).
Moreover sometimes an experimentalist has a limited time for doing the measurement
and he is not able to accumulate as many statistical data as it would be necessary.
One of the trivial reason for that might be instability of the experimental set-up.
So practically any measurement set has a limited size and it is not convenient
(or even is not possible) to increase it for achieving complete volume. That is why
it would be useful to take into account the set of available tools and develop a
protocol which gives the highest accuracy for fixed experimental resources.\\
The present work is addressed to experimental problem of realization
of the optimal state reconstruction for biphoton-based polarization
ququarts.  We restrict ourselves with the protocol of quantum
tomography suggested and tested earlier \cite{ourPRA:06}. Basically
the method is based on an optimal choice of the measuring scheme's
parameters that provides better quality of reconstruction with the
fixed set of statistical data.

\section{Quantum tomography. Principle of realization.}
An arbitrary quantum state is completely determined by a wave
vector for pure state, or by a density matrix for mixed state. To
measure the quantum state one needs to perform a set of projective
measurements and then to apply some computation
procedure to the data obtained at the previous stage. It is well known that the number of real
parameters characterizing a quantum state is determined by the
dimension of the Hilbert space $d$. For a pure state,
\begin{equation}
N_{pure}=2d-2, \label{eq:pure}
\end{equation}
end for mixed state,
\begin{equation}
N_{mixed}=d^2-1. \label{eq:mixed}
\end{equation}
However in practice the normalization is necessary to be established
with the data so the total number of measurements increases by one.
According to Boht's complementarity principle, it is impossible to
measure all projections simultaneously, operating with single
quantum state only. So, first of all, one needs to
generate a lot of the same representatives of a quantum ensemble \cite{D'Ariano}. In our
experiment for preparation such states we used the process of
spontaneous parametric down-conversion (SPDC). So fixing the conditions
under which SPDC takes place we achieve the initial states which serve as a base for further manipulations.\\
As it was already mentioned above we deal with the collinear and
frequency non-degenerate regime of biphoton field for which
$k_{s}|| k_{i}$, and $w_{s}\ne w_{i}$.  From the point view of polarization there are four
natural states of photons pairs: $|H_{s}H_{i}>, |H_{s}V_{i}>,
|V_{s}H_{i}>, |V_{s}V_{i}>$. Then any pure polarization state of
biphoton can be expressed as superposition of four basis
states:
\begin{equation}
|C>=c_{1}|H_{s}H_{i}>+c_{2}|H_{s}V_{i}>+c_{3}|V_{s}H_{i}>+c_{4}|V_{s}V_{i}>.
\label{eq:state}
\end{equation}
Here $c_i=|c_i|e^{i\phi_i}$, $\sum\limits_{i = 1}^4 {| {c_i} |^2 =
1} $ are complex probability amplitudes. Thus ququart represents a
quantum polarization state of two qubits (photons), whose states can
be either entangled or non-entangled. For complete characterization
of polarization ququarts and their properties including methods for
preparation, transformation and measurement we refer to our previous work \cite{ourPRA:06}.\\
It is worthy to note that the universally accepted method for
describing the multi-mode quantum polarization states of photons is
based on P-quasispin approach \cite{Karassiov93}. Application of
P-quasispin concept to the polarization ququart has been done in
\cite{KaraKul07}. In paper \cite{dnk:97} has been shown, that
polarization properties of two-mode biphoton field are completely
defined by the coherency matrix. It is a matrix consisting of
fourth-order moments in the electromagnetic field
\begin{equation}
K_4 =\left( {{\begin{array}{*{20}c}
 A \hfill & E \hfill & F \hfill & G \hfill\\
 {E^ * } \hfill & B \hfill & I \hfill & K \hfill\\
 {F^ * } \hfill & {I^ * } \hfill & C \hfill & L \hfill\\
 {G^ * } \hfill & {K^ * } \hfill & {L^ *} \hfill & D \hfill\\
\end{array} }} \right),
\label{eq:K4}
\end{equation}
\begin{equation}
{{\begin{array}{*{20}c} A \equiv \langle
a_{s}^{\dagger}a_{i}^{\dagger}a_{s}a_{i}\rangle=|{c_1}|^2, \quad
 B \equiv \langle
a_{s}^{\dagger}b_{i}^{\dagger}a_{s}b_{i}\rangle=|{c_2}|^2,\hfill &\\
 C \equiv \langle
b_{s}^{\dagger}a_{i}^{\dagger}b_{s}a_{i}\rangle=|{c_3}|^2, \quad D
\equiv \langle
b_{s}^{\dagger}b_{i}^{\dagger}b_{s}b_{i}\rangle=|{c_4}|^2.\hfill &\\
 E \equiv \langle
a_{s}^{\dagger}a_{i}^{\dagger}a_{s}b_{i}\rangle=c_1^*c_2, \quad
 F \equiv \langle
a_{s}^{\dagger}a_{i}^{\dagger}b_{s}a_{i}\rangle=c_1^*c_3,\hfill &\\
 G \equiv \langle a_{s}^{\dagger}a_{i}^{\dagger}b_{s}b_{i}\rangle=c_1^*c_4,
\quad I \equiv \langle a_{s}^{\dagger}b_{i}^{\dagger}b_{s}a_{i}\rangle=c_2^*c_3,\hfill &\\
K \equiv \langle
a_{s}^{\dagger}b_{i}^{\dagger}b_{s}b_{i}\rangle=c_2^*c_4,
\quad L \equiv \langle b_{s}^{\dagger}a_{i}^{\dagger}b_{s}b_{i}\rangle=c_3^*c_4.\hfill &\\
\end{array} }}
\label{eq:elements}
\end{equation}
The averaging in (\ref{eq:elements}) is taken over the state
(\ref{eq:state}). The polarization density matrix of ququart state
coincides with coherency matrix $K_4$ and completely determines an
arbitrary ququart state. Thus in order to reconstruct the
unknown ququart state all moments (\ref{eq:elements}) has to be measured\\
As it was mentioned earlier, to measure an unknown state it is
necessary to perform a set of projective measurements. At present,
the only realistic way to register fourth-order moments is using the
Hanbury Brown-Twiss scheme. In order to be able to measure
polarization moments (\ref{eq:elements}) we supplied this scheme
with retardant plates and polarization prisms. Basically two
protocols for quantum state reconstruction of ququarts can be
applied \cite{ourPRA:06}. In the first protocol the ququart is
divided into two spacial/frequency modes by means of dichroic mirror
and then each photon of the pair is subjected by polarization
transformations separately. In the second protocol the ququart
undergoes polarization (linear) transformations as a whole before
the beamsplitter. Here we consider only the second protocol since it
seems to be more practical. The idea of the protocol is
straightforward. The ququart state is transformed by two retardant
plates Wp1, Wp2 putting in series and after that it is split by
beamsplitter into two spatial modes ended with single-photon
detectors D (Fig.2.1). Polarization prisms project the state onto
vertical polarization. Pulses coming from detectors are coupled on
the coincidence scheme, which selects only those of them coinciding
in time with the accuracy of coincidence window (about 3 nsec in our
case). So each pulse coming from the coincidence scheme associates
with projection of the initial ququart subjected to given
polarization transformation. Finally the output pulses of the
coincidence scheme accumulated during fixed time interval
(coincidence rate) serve as statistical data to be analyzed for the
state reconstruction.
\begin{figure}[!dt]\center
\includegraphics[width=0.9\textwidth]{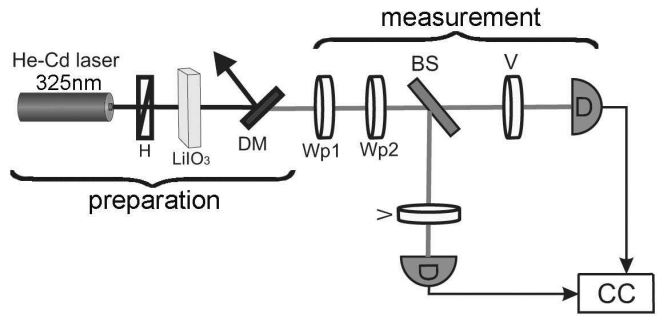}
\caption{Setup for preparation and measurement of ququarts}
\end{figure}
The transformations performed by the plates Wp1, Wp2 are expressed by form:
\begin{equation}
|\Psi^{out}\rangle_{kl}=\hat{G}(\delta_{1(s,i)},\theta_{k})\hat{G}(\delta_{2(s,i)},\theta_{l})|\Psi^{in}\rangle.
 \label{eq:protocol1}
\end{equation}
Matrix $G$ is given by $4\times 4$ matrix which is obtained by
a direct product of two $2\times 2$ matrices describing the $SU(2)$
transformation performed on each photon \cite{dnk:97}:
\begin{equation}\footnotesize
\hat{G} \equiv \left( {{\begin{array}{*{20}c}
 {t_{s}t_{i}} \hfill & {t_{s}r_{i}} \hfill & {r_{s}t_{i}} \hfill & {r_{s}r_{i}} \hfill \\
 {-t_{s}r_{i}^*} \hfill & {t_{s}t_{i}^*} \hfill & {-r_{s}r_{i}^*} \hfill & {r_{s}t_{i}^*} \hfill \\
 {-r_{s}^*t_{i}} \hfill & {-r_{s}^*r_{i}} \hfill & {t_{s}^*t_{i}} \hfill & {t_{s}^*r_{i}} \hfill \\
 {r_{s}^*r_{i}^*} \hfill & {-r_{s}^*t_{i}^*} \hfill & {-t_{s}^*r_{i}^*} \hfill & {t_{s}^*t_{i}^*} \hfill \\
\end{array} }} \right) =
\left({{\begin{array}{*{20}c}
 {t_{s}} \hfill & {r_{s}} \hfill \\
 {-r_{s}^*} \hfill & {t_{s}^*} \hfill \\
\end{array} }} \right)\otimes
\left({{\begin{array}{*{20}c}
 {t_{i}} \hfill & {r_{i}} \hfill \\
 {-r_{i}^*} \hfill & {t_{i}^*} \hfill \\
\end{array} }} \right),
\label{eq:gmatrix}
\end{equation}
with complex coefficients of effective transmission
\begin{equation}
t_{1,2(s,i)}=\cos\delta_{{1,2}(s,i)}+i\sin\delta_{{1,2}(s,i)}\cos2\theta_{k,l},
\label{eq:transmit}
\end{equation}
and effective reflection
\begin{equation}
r_{1,2(s,i)}=i\sin\delta_{{1,2}(s,i)}\sin2\theta_{k,l}.
\label{eq:reflect}
\end{equation}
Here $\theta_{k,l}$ are the orientation angles of the first or
second retardant plates. The parameters of the plates, i.e. optical
thicknesses for different wavelengths $\delta_{{1,2}(s,i)}$ and
their orientations, are supposed to be known with high accuracy
which relates to the final accuracy of the state reconstruction.
Disregarding the normalization, the number of events detected in the
experiment, i.e. coincidence rate $R_{kl}$ is the projection of the
transformed state $|\Psi_{out}\rangle$ onto the state
$|V_{s}V_{i}\rangle$ determined by the orientation of the
polarization prisms. This projection is given by the expression
\begin{equation}
R_{kl}\propto|\langle V_{s}V_{i}|\Psi_{out}\rangle_{kl}|^2.
 \label{eq:coinc1}
\end{equation}
Thus the joint action of two retardant plates and polarization
prisms provides the basis for projective measurements.\\
The intensity of the event generation in each process can be
expressed in terms of squared modulus of the amplitude of a quantum
process \cite{root-est}
\begin{equation}
R_{\nu}=M_{\nu}{^\star}M_{\nu}, \label{eq:moments}
\end{equation}
Although the amplitudes of the processes cannot be measured
directly, they are of the greatest interest as these quantities
describe fundamental relationships in quantum physics. It follows
from the superposition principle that the amplitudes are linearly
related to the state-vector components \cite{root-est}. So the main purpose of
quantum tomography is the reproduction of the amplitudes and state
vectors, which are hidden from direct observation. The linear
transformation of the state vector $c =(c_{1},c_{2},c_{3},c_{4})$
into the amplitude of the process M is described by a certain matrix
$X$
\begin{equation}
Xc=M. \label{eq:instrumenntalequation}
\end{equation}
The matrix $X$ is a so called instrumental matrix for a set of
mutually complementary measurements. Suppose that protocol contains
$m$ steps, which means that $m$ consistent transformations should be
done with the initial state. Consequently the matrix $X$ contains
$m$ rows. So each row corresponds to various projection measurements
under a quantum state. To each row $X_{j}$ $j=1,2,...,m$ of a length
$d$ we will construct a new row of a length $d^{2}$ which is
determined as a direct product of row $X_{j}$ and row $X_{j}^\star$.
Then let us compose another matrix $B$ with rows introduced above.
The size of the matrix $B$ is $m\times d^2$ and we assume that $m
\geq d^2$. The important property of the protocol of measurement is
its completeness. The protocol is supposed to be complete if all
$d^2$ singular eigenvalues of the matrix $B$ are strictly positive.
Such protocol provides with a reconstruction of an arbitrary quantum
state at sufficiently large sample size. However if several (or even
single) singular eigenvalues of the matrix are close to zero then
the matrix becomes degenerate.  Hence in order to reconstruct
unknown ququart state with this quasi-degenerate matrix one needs to
increase number of statistical data. We have found rank and singular
eigenvalue of the matrix $B$ and introduced parameter R which is
defined as the ratio between the minimal nonzero singular eigenvalue
and maximal one. The R value lies in a range from 0 to 1. It is
important to notice that the ratio R depends on parameters of
experimental set-up only (in out case these parameters are optical
thicknesses of the quartz plates) and does not depend at all on the
state to be measured. Therefore the ratio R can serve as a testing
parameter of the protocol in the sense whether the protocol is
optimal or not. Namely the smaller R the worse the protocol and the
quality of the reconstructed state is supposed to be lower. In the
present paper we are not going to prove this statement
mathematically. We just suggest the empirical parameter and test its
validity with particular experiments. In nearest future we will
develop this concept and give strict arguments related to the
completeness of statistical state reconstruction protocol \cite{xxx}
. In our experiment each run is specified by the orientation angle
of the Wp1 plate $\theta_{1}=0^\circ,15^\circ,30^\circ,45^\circ$ for
the complete rotation of the Wp2 plate by $360^\circ$ with step
$10^\circ$; i.e., 144 measurements have totally been made.
Consequently in our case matrix $X$ consists of 144 rows (the total
number of different orientations for both plates in experiment) and
4 columns (the dimension of Hilbert space for ququarts). Each row is
formed in the following way. The initial state $|\Psi_{in}\rangle$
is transformed by the two quarts retardant plates Wp1 and Wp2 and
projected onto the vertical state $|V_{1}V_{2}\rangle$. Thus using
the formulas (\ref{eq:gmatrix}-\ref{eq:reflect}) the four-element
row can be re-written in the form
\begin{equation}
\begin{array}{*{20}c}
X_{j}=1/2(\alpha_{s}\alpha_{i}& \alpha_{s}\beta_{i}&
\beta_{s}\alpha_{i}& \beta_{s}\beta_{i})
\end{array}
\end{equation}
where the values of the complex parameters $a_{s,i}, b_{s,i}$ are
different for particular rows of the measurement protocol:
\begin{equation}
\begin{array}{cc}
\alpha_{s,i}=-t_{1(s,i)}^\star(\theta_{k})r_{2(s,i)}(\theta_{l})-r_{1(s,i)}(\theta_{k})t_{2(s,i)}(\theta_{l})\\
\beta_{s,i}=-r_{1(s,i)}^\star(\theta_{k})r_{2(s,i)}(\theta_{l})+t_{1(s,i)}(\theta_{k})t_{2(s,i)}(\theta_{l}).
\label{eq:instrumenelements}
\end{array}
\end{equation}
Here indexes $1,2$ relate to the first(Wp1) or second(Wp2) retardant
plates correspondingly. Within the bounds of the proposed method of
measurement thicknesses of plates would seem does not play a
significant role and can be chosen arbitrary. Of course the
experimentalist should know the exact optical thickness of each
plate to be used in further calculations. However, below (see
Sec.III) we will shown, that only particular sets of plates with
certain thicknesses provide the optimal reconstruction of an unknown
state.

\section{Quantum tomography. Simulation and experiment.}

Let us analyze how the choice of plates's thicknesses affect on the
quality of state reconstruction. We have calculated ratio R for set
of thicknesses of the first and second quartz retardant plates.
Specifically first plate was varied within the limits from 0.8mm to
1mm with the step 0.002mm and second plate was varied within the
limits from 0.5mm to 1mm with the same step. Corresponding picture
is shown on Fig.3.1. Here along the axis $x$, $y$ lie thicknesses of
the first and second quarts retardant plates correspondingly.

\setcounter{figure}{0}

\begin{figure}[!hb]\center
\includegraphics[width=0.8\textwidth]{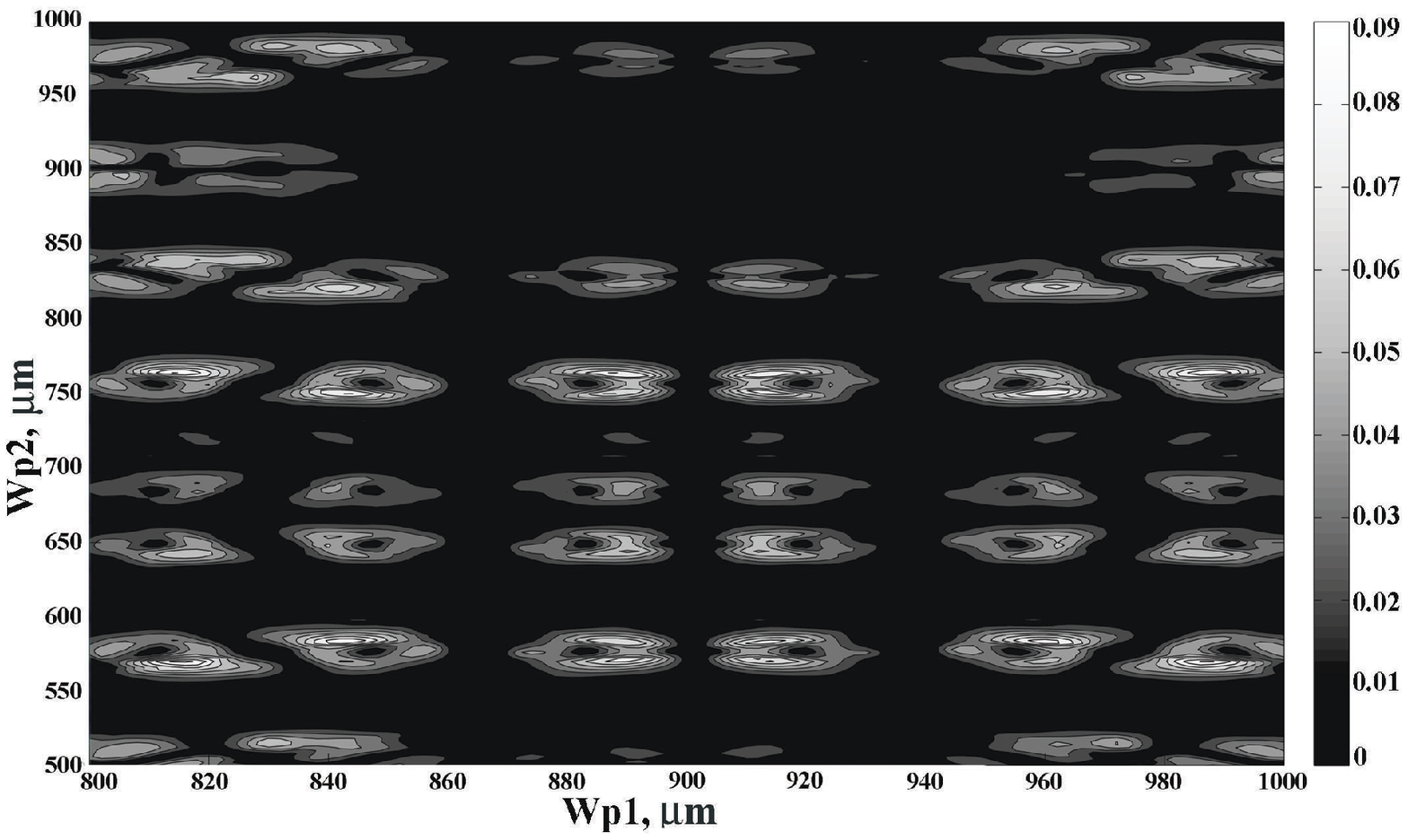}
\caption{Dependence of ratio R on plates thicknesses}
\end{figure}

On Fig.3.1 light color indicates areas with high value of ratio R,
and dark color indicates areas with low value. It is obvious, that
the rash choice of plates thicknesses most likely will be wrong from
the point of view of completeness of matrix B. For offered protocol
the maximum achievable parameter R takes on a value 0.09 and as it
will be shown below, it is quite enough for good quantum state
reconstruction. For example, to receive such value one can choose
plates with thickness Wp1=0.988mm, Wp2=0.570mm. Particularly for our
experiment we have chosen the following two sets of plates: a)
Wp1=0.988mm,Wp2=0.836mm (optimal) ; b) Wp1=0.836mm, Wp2=0.536mm
(non-optimal). For first set of plates the ratio is equal to
$R=0.0554$, for second set $R=0.0013$. Unfortunately, our choice has
been limited by available plates, so we have not taken the extremal values.\\
In the simplest case, for example for state
$|V_{s}\rangle|V_{i}\rangle $ we can simulate the procedure of the
basis state reconstruction. The difference between reconstructed
sate (which was numerically simulated) and theoretical one is caused
by information loss due to the finite number of statistical data. As
estimated parameters, we have considered the  average information
losses
\begin{equation}
L=log_{10}\frac{1}{1-\bar{F}},\label{eq:losses}
\end{equation}
where $F(fidelity)$ represents a correspondence between the
theoretical and experimental (or numerically simulated) state
vectors at different thicknesses of the first and second quarts
retardant plates. Corresponding plot is shown on Fig.3.2. Here the
thicknesses of the first and second quarts retardant plates lie
along the axis $x$, $y$. Thicknesses of both plates were varied
within the limits indicated on previous plot.
\begin{center}
\begin{figure}[!dt]\center
\includegraphics[width=0.8\textwidth]{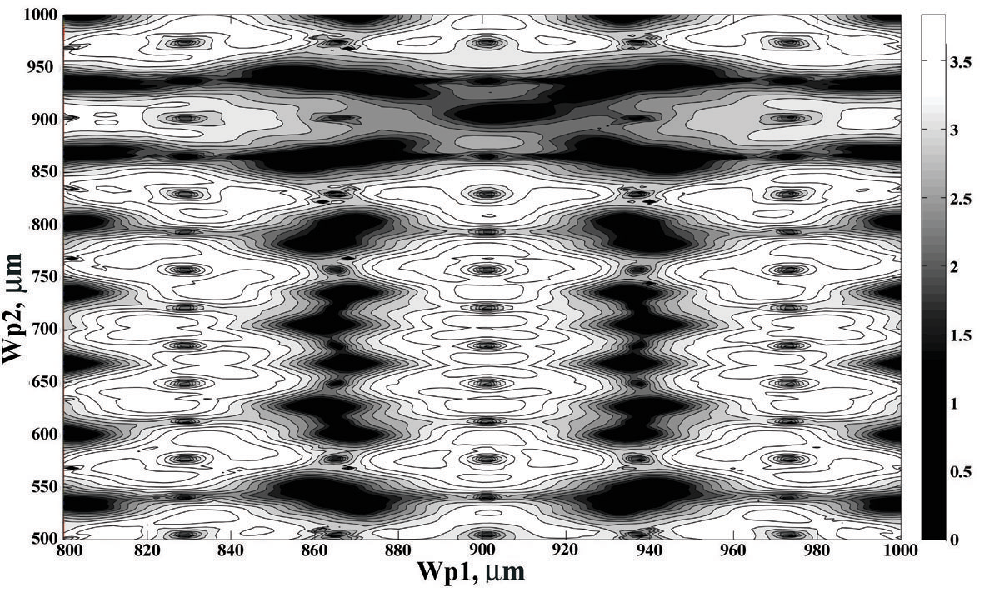}
\caption{Dependence of average information losses $L$ on plates
thicknesses}
\end{figure}
\end{center}
On Fig.3.2 light color indicates areas with low value of information
losses, and dark color indicates areas with high value of
information losses. Numbers at the right bar show the accuracy level
of the tomography protocol. For example, value $3$ means, that
correspondence between the theoretical and experimental state
vectors ($fidelity$) is about 0.999. "Black holes" on Fig.3.2 set
areas, where matrix $B$ becomes ill-conditioned. In other words the
figure presented at the plot serves as some sort of a "navigation
map" for measurement protocol. To achieve good quality of the
measurement one should select the plates thicknesses in light areas
and avoid dark areas. It is clearly seen that minima (dark areas) on
both plots (Fig.3.1 and Fig.3.2) coincide. Moreover Fig.3.1 shows
that choosing the plates one mets more limitations since the ratio R
does not depend on the state to be measured. At the same time the
parameter "average information losses $L$" does. This fact has to be
taken into account by performing statistical state reconstruction.
The dependence of density of distribution on the accuracy
information losses $1-F$ is shown in Fig.3.3. We see that accuracy
information losses for the optimal tomography are characterized by
density of distribution which is determined by the narrow high peak
concentrated at low values of information losses (obviously more low
0.005). On the contrary, non-optimal tomography is characterized by
the wide density of distribution stretched up to values 0.05. These
differences disappear at sufficiently large sample size, but for
equal quality of state reconstruction the right choice of plates
allows one to make smaller set of statistical data, i.e. reduce time
exposition.\
\begin{figure}[!dt]\center
\includegraphics[width=0.7\textwidth]{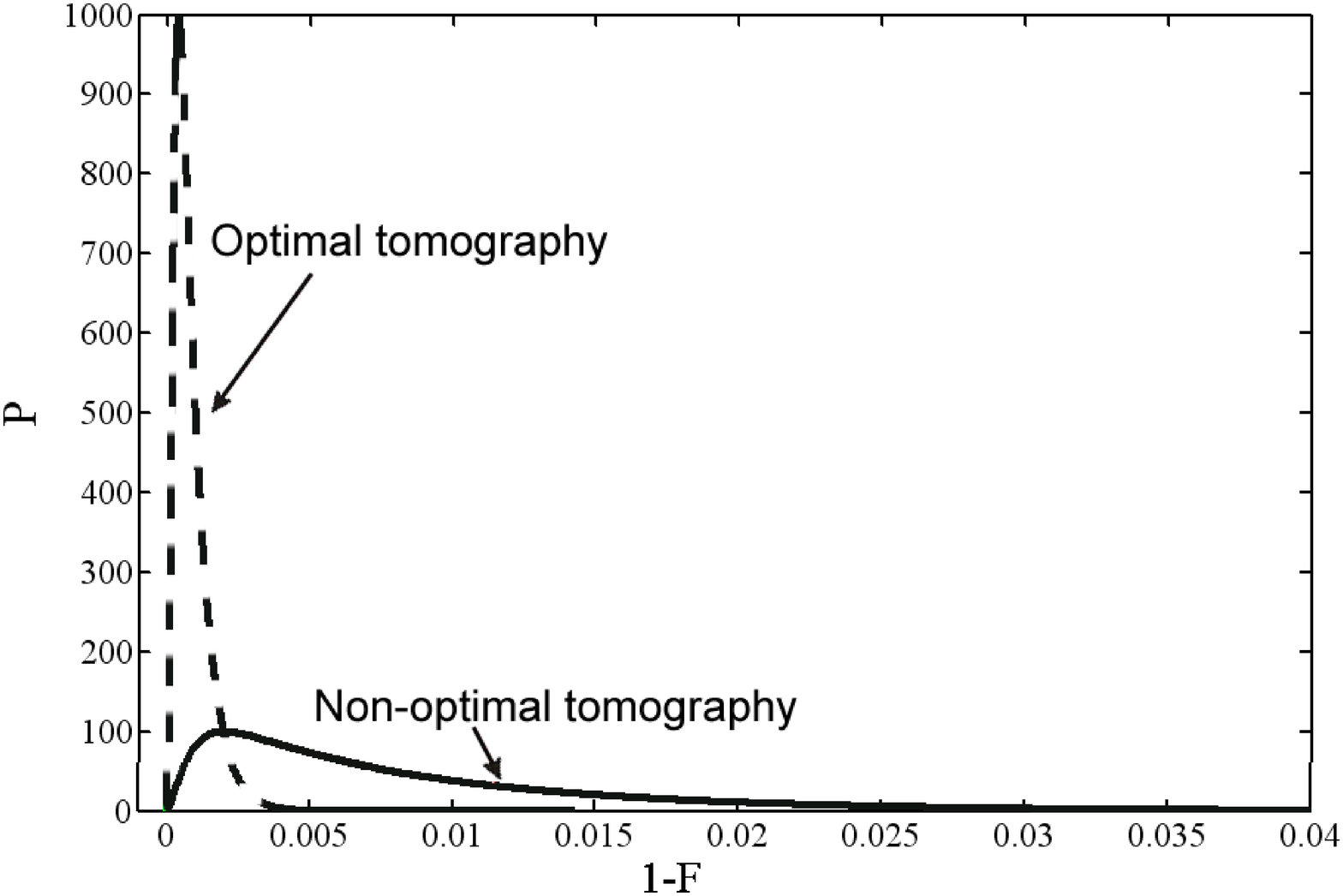}
\caption{Dependence of density of distribution on the accuracy
information losses $1-F$.  Dot line corresponds to optimal set of
plates (optimal tomography), solid line corresponds to non-optimal
set (non-optimal tomography.)}
\end{figure}\\
We applied suggested protocol to measure some particular set of
ququart states $\Psi$. For the generation of biphoton-based ququarts
we used lithium-iodate 15 mm crystal (with type-I phase matching)
pumped with 5 mW cw horizontally polarized helium-cadmium laser
operating at 325 nm. The angle between the pumping wavevector and
optical axis of the crystal is equal to $58^\circ$. Under these
conditions, the state $|V_{702nm}V_{605nm}\rangle$ is generated in
the crystal. The initial state was subjected to transformations done
by dichroic retardant plate in order to prepare some subset of
ququart. This subset is known as a product of two polarization
qubits \cite{product}. That subset of states was used to be
reconstructed. In particular, we used the 0.441mm length quartz
plate and changed it orientation. We tested the states which were
generated for three orientation angles of plate QP $\alpha= 0^\circ,
-60^\circ$. Since the thickness of the plate, quartz dispersion and
orientation were supposed to be known, we were able to calculate the
result of the state transformation with high accuracy. In
measurement part of setup we used two sets of retardant plates with
different thicknesses, optimal and non-optimal, and made the
reconstruction procedure of ququarts states at the fixed set of
statistical data for each set. In both series of experiments we
gathered identical statistic equal 30-35 thousands events. According
to protocol, four sets of measurements were performed for each input
state, so totally we performed 144 measurements of the coincidence
rate as a function of orientation angles of Wp1 and Wp2. When
processing experimental data, the random coincidence rate $Nran$,
which is expressed in terms of the rate of averaged single counts
from each photodetector and the coincidence-window width ($T$) of
the scheme as $N_{ran} = \langle N_{1}\rangle \langle N_{2}\rangle
T$, is extracted from the coincidence number. For state
reconstruction we used the maximum likelihood method, that was
developed in \cite{Bogdanov} and was successively applied to
reconstruct states of optical qutrits \cite{ourPRA:04}. The result
of ququarts reconstruction is given in Table 1.
\begin{table}[!ht]\center
\caption{ }
\begin{tabular}{|c|c|c|c|c|c|} \hline
\multicolumn{1}{|c|}{} & \multicolumn{2}{c|}{fidelity} \\
\hline $\alpha$ (deg.) & {optimal set} & {non-optimal set}\\
\hline   0 &  0.999 & 0.974 \\
\hline -60 &  0.993 & 0.975 \\
\hline
\end{tabular}
\end{table}
The value of parameter F is defined as $F=|\langle c_{theory}|
c_{exp}\rangle|^2$. It is clearly see that obtained fidelity values
for optimal set of thicknesses is much higher than for non-optimal
set. Fidelity for all states in the first case (optimal) was above
0.99, in the second case (non-optimal) fidelity was about 0.97.\\
\section{Conclusion}In conclusion, we have suggested and tested a sort of
an optimal protocol for polarization ququarts
state tomography. The protocol allows one to achieve highest
accuracy of the state reconstruction with available resources which
experimentalist holds in his hands while doing particular
measurements. Then we investigated theoretically and experimentally
the accuracy of polarization ququart reconstruction depending on
parameters available in experiment (quartz plates thicknesses) at
fixed set of statistical data. The developed methodology can be extended easily
to other quantum states reconstruction as well as used for optimization of various
technological parameters of quantum tomography protocols.\\
\section{Acknowledgments} This work was supported, in part, by Russian Foundation for Basic Research
(Projects 06-02-16769 and 06-02-39015) and the Leading
Russian Scientific Schools (Project 796.2008.2). E.V.M. acknowledges the support of the Dynasty
Foundation.


\begin{thebibliography}{12}
\bibitem{product} Since we started with the product state $|V_{702nm}V_{605nm}\rangle$ its zero entanglement degree remains constant under local polarization transformations
\end{thebibliography}
\end{document}